\documentclass[%
 reprint,
 prl,
 floatfix,
 amsmath,amssymb,
 aps, widetext, superscriptaddress
]{revtex4-2}

\usepackage{xcolor}
\usepackage{hyperref}
\usepackage{graphicx}
\usepackage{dcolumn}
\usepackage{bm}
\usepackage{microtype}
\usepackage{physics}
\usepackage{xspace}
\usepackage{chemformula}
\usepackage{csquotes}
\usepackage{lipsum}
\usepackage{comment}
\usepackage{todonotes}

\newcommand{\hl}[1]{{#1}}

\newcommand{\rv}{\mathbf{r}}
\newcommand{\Ev}{\mathbf{E}}
\newcommand{\Jv}{\mathbf{J}}

\allowdisplaybreaks[3]

\newcommand{\tuwien}{Institute for Theoretical Physics, Vienna University of Technology (TU Wien), Vienna A-1040, Austria}

\begin{document}
\title{Geometric Pumping of Optical Work in Coherent Wave Scattering}

\author{Zeyu Kuang}
\thanks{zeyu.kuang@tuwien.ac.at}
\affiliation{\tuwien}

\author{Konstantin Y. Bliokh}
\affiliation{Donostia International Physics Center (DIPC), Donostia-San Sebastián 20018, Spain}
\affiliation{IKERBASQUE, Basque Foundation for Science, Bilbao 48009, Spain}
%\affiliation{Centre of Excellence ENSEMBLE3 Sp. z o.o., 01-919 Warsaw, Poland}

\author{Stefan Rotter}
\affiliation{\tuwien}

\date{\today}

\begin{abstract}
%Adiabatic cyclic modulation of two or more parameters can pump electrons through a system without external bias, a foundational mechanism in mesoscopic transport. Here we show that an analogous principle governs energy exchange in coherent electromagnetic scattering: a finite amount of optical work can be pumped between light and an external drive by a cyclic modulation of two or more parameters in the scattering potential. Unlike electron pumping between incoherent reservoirs, our optical analog relies on fully coherent, structured incident fields. As a consequence, completely incoherent illumination pumps no net optical work, whereas coherent illumination enables both the generation and optimization of pumped work through wavefront shaping. In particular, we derive the optimal wavefront and optimal modulation trajectory that maximize the pumped optical work per cycle. Our theory connects time-varying photonics, opto-mechanics, and geometric transport within a common scattering framework, and can be used to design efficient nano-motors and optical frequency modulators.  
Optical work exchanged between light and time-varying matter underlies a wide range of emerging photonic technologies, yet a general geometric theory describing its generation and control has been lacking. Here we establish such a theory for coherent electromagnetic scattering. We show that the net optical work accumulated during an adiabatic modulation cycle is a geometric quantity determined by the Berry curvature of the \hl{outgoing field} in parameter space.
\hl{Nonzero optical work can thus be generated by cyclically modulating two or more independent system parameters. In contrast to conventional adiabatic pumping between incoherent reservoirs, optical work pumping vanishes under maximally mixed illumination. Greater illumination coherence increases the maximum work attainable through wavefront shaping. For fully coherent illumination,} we derive the optimal incident wavefront and modulation trajectory that maximize the optical work per cycle. Our framework unifies time-varying photonics, optomechanics, and geometric transport within a common scattering formalism. %, providing practical design principles for light-driven nano-motors, optical frequency conversion, and other time-varying photonic devices.
\end{abstract}

\maketitle

{\it Introduction.---}%Adiabatic charge pumping is a paradigmatic mechanism of driven transport: by slowly and cyclically modulating two or more parameters of a scattering potential, currents can be driven through a system even in the absence of any external bias~\cite{thouless_quantization_1983, citro_thouless_2023,brouwerScatteringApproachParametric1998,switkesAdiabaticQuantumElectron1999a}. In mesoscopic electron transport, Brouwer's scattering theory of parametric pumping~\cite{brouwerScatteringApproachParametric1998} established that while a single adiabatically varying parameter does not pump any net charge over a cycle, the simultaneous modulation of two or more parameters does. This mechanism has since become a unifying paradigm across spin pumping in spintronics~\cite{sharmaQuantumPumpSpin2001, muccioloAdiabaticQuantumPump2002, tserkovnyakEnhancedGilbertDamping2002, tserkovnyakSpinPumpingMagnetization2002,tserkovnyakNonlocalMagnetizationDynamics2005,chengSpinPumpingSpinTransfer2014}, Brownian motors~\cite{purcellLifeLowReynolds2012, astumianBrownianMotors2002,hanggiArtificialBrownianMotors2009,sinitsynStochasticPumpEffect2009}, and heat pumping through electrons~\cite{wangHeatCurrentParametric2002, avronTransportDissipationQuantum2004,arracheaHeatProductionEnergy2007,reyNonadiabaticElectronHeat2007}, photons~\cite{liAdiabaticThermalRadiation2019, fernandez-alcazarImplementationOptimalThermal2021, fernandez-alcazarExtremeNonreciprocalNearField2021}, or quantum two-level systems~\cite{segalMolecularHeatPump2006, ludovicoPeriodicEnergyTransport2016,bhandariGeometricPropertiesAdiabatic2020, arracheaEnergyDynamicsHeat2023}. Most of these studies, like the original work of Brouwer, concern the transfer of conserved quantities between thermal reservoirs, that is, between incoherent sources.
Adiabatic pumping has become a paradigmatic mechanism of driven transport, in which slowly and cyclically modulating 
%two or more 
parameters of a scattering potential generates a net transport of conserved quantities even in the absence of any external bias~\cite{thouless_quantization_1983, citro_thouless_2023,brouwerScatteringApproachParametric1998,switkesAdiabaticQuantumElectron1999a}. In mesoscopic electron transport, Brouwer's scattering theory of parametric pumping~\cite{brouwerScatteringApproachParametric1998} established that while a single adiabatically varying parameter does not pump any net charge over a cycle, the simultaneous modulation of two or more parameters does. This geometric mechanism has since found applications across spin pumping in spintronics~\cite{sharmaQuantumPumpSpin2001, muccioloAdiabaticQuantumPump2002, tserkovnyakEnhancedGilbertDamping2002, tserkovnyakSpinPumpingMagnetization2002,tserkovnyakNonlocalMagnetizationDynamics2005,chengSpinPumpingSpinTransfer2014}, Brownian motors~\cite{purcellLifeLowReynolds2012, astumianBrownianMotors2002,hanggiArtificialBrownianMotors2009,sinitsynStochasticPumpEffect2009}, \hl{quantum adiabatic motors~\cite{bodeScatteringTheoryCurrentInduced2011, bustos-marunAdiabaticQuantumMotors2013}}, and heat pumping through electrons~\cite{wangHeatCurrentParametric2002, avronTransportDissipationQuantum2004,arracheaHeatProductionEnergy2007,reyNonadiabaticElectronHeat2007}, photons~\cite{liAdiabaticThermalRadiation2019, fernandez-alcazarImplementationOptimalThermal2021, fernandez-alcazarExtremeNonreciprocalNearField2021}, and quantum two-level systems~\cite{segalMolecularHeatPump2006, ludovicoPeriodicEnergyTransport2016,bhandariGeometricPropertiesAdiabatic2020, arracheaEnergyDynamicsHeat2023}. All of these studies, however, concern the transport of conserved quantities between incoherent reservoirs.
%thermal reservoirs, that is, between incoherent sources.

\begin{figure}[b!]
    \centering
    \includegraphics[width=1\linewidth]{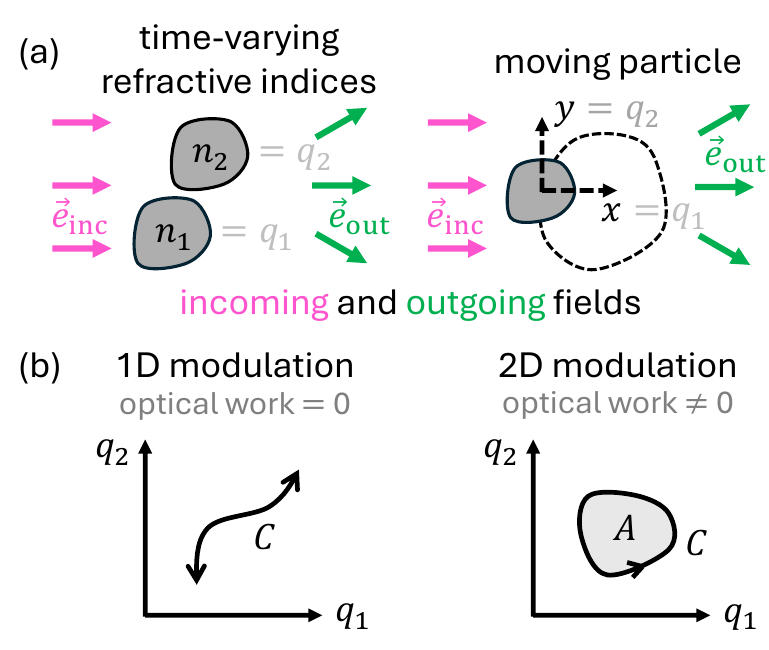}
    \caption{(a) Light scattering in two representative time-varying systems: varying refractive indices $(n_1,n_2)$ (left), and a particle with varying coordinates $(x,y)$ in a 2D plane (right). 
    (b) The two 
    %control 
    parameters $(q_1,q_2)$ can be modulated either along a 1D path (left panel) or along a closed 2D loop $C$ in parameter space (right panel). The latter encloses an area $A$ and thereby generates a finite net optical work.} 
    \label{fig:1}
\end{figure}

\hl{Here we extend the theory of adiabatic pumping from incoherent to coherent sources, focusing on the important case of coherent electromagnetic scattering where the pumped quantity is optical work.}
%Here we establish a geometric theory of optical work in coherent electromagnetic scattering. 
We show that a cyclic modulation of two or more scattering parameters, such as refractive indices of tunable media or coordinates of moving particles [Fig.~\ref{fig:1}(a)], generally transfers a finite amount of energy between light and an external drive through a geometric mechanism that shares the same mathematical structure as adiabatic charge pumping in mesoscopic transport. Consequently, while a one-dimensional (1D) adiabatic modulation in parameter space does not produce any net optical work over a cycle, a two-or-more-dimensional adiabatic modulation generally does [Fig.~\ref{fig:1}(b)]. Our theory reveals optical work as a geometric quantity determined by the Berry connection and curvature of the \hl{outgoing field} in parameter space. Unlike conventional adiabatic pumping between incoherent reservoirs, 
%however, 
our mechanism \hl{benefits from} the coherence in the \hl{illumination}: incoherent illumination \hl{where all input channels are equally and incoherently populated} pumps no net optical work, whereas coherent illumination enables finite work generation, \hl{with more coherence (measured in terms of majorization order~\cite{guoTransportMeasurementsMajorization2025}) consistently leading to more work} after wavefront shaping. In particular, both the incident wavefront and the modulation trajectory can be optimized to maximize or minimize the pumped optical work based solely on the instantaneous scattering matrix.

In the following, we prove these results and numerically test them in two representative examples: (i) 1D light scattering from dielectric slabs with tunable permittivities, and (ii) 3D light scattering from a moving particle. These examples illustrate the generality of our theory and its potential applications in time-varying photonics~\cite{galiffi_photonics_2022} and optomechanics~\cite{aspelmeyerCavityOptomechanics2014a, toftulRadiationForcesTorques2026}.

{\it General formalism.---} We consider a time-varying scattering system, such as those shown in Fig.~\ref{fig:1}(a). The system is governed by a set of time-varying parameters $\bm{q}(t) = \left( q_1(t), q_2(t), \dots, q_N(t) \right)$ ($N=2$ in Fig.~\ref{fig:1}).
%, which could be, for example, the permittivities of electrically tuned media or the coordinates of moving particles. 
%Shown in Fig.~\ref{fig:1}(a) are two examples with $N=2$.
We assume that the scattering medium is linear and lossless (i.e., no material absorption), and light comes in monochromatically at frequency $\omega$. We represent the incoming light field by an $M$-component vector $\vec{e}_{\rm inc}$ in an orthonormal basis, such as spherical waves in free space or waveguide modes in a waveguide, \hl{normalized by their power flux}.   
Upon scattering, light transfers energy to the time-varying system in the form of optical work \hl{$W$}, and leaves the system at some Doppler-shifted frequencies. 
\hl{Quantitatively, this optical work has the rate: $dW/dt = \left\{\int_V |\bm{E}(\bm{r},t)|^2 d\varepsilon[\bm{r},\bm{q}(t)]\right\}/dt$ where $d\varepsilon[\bm{r},\bm{q}(t)]$ is the change of the permittivity under the control parameter $\bm{q}(t)$ at time $t$ and position  $\bm{r}$ (assuming a non-magnetic material), $\bm{E}(\bm{r},t)$ is the electric field at those points, and $V$ denotes the whole space [see Supplementary Material (SM)~\cite{SM} for a detailed discussion].} The goal of this Letter is to quantify this optical work in terms of the trajectory of $\bm{q}(t)$ in parameter space [see Fig.~\ref{fig:1}(b)], and to understand when this work can be maximized (with either positive or negative sign) or reduced to zero. 

%A central task in time-varying photonics is to study and maximize this energy transfer by tuning the system parameters $\bm{q}(t)$ and the incoming field $\vec{e}_{\rm inc}$.

We focus on the limit where the parameter $\bm{q}(t)$ changes adiabatically in time, which is the regime most electromagnetic experiments operate in.
Correspondingly, 
%we assume that 
the typical inverse time scale of the parameter variation, $|{q_i}^{-1}d{q_i} / dt|$ is assumed to be much smaller than the light frequency $\omega$ and the relaxation rates of the resonant modes in the system. Then, light scattering in the system can be described by the instantaneous scattering matrix $\hat{S}(\bm{q})$ acting on $\vec{e}_{\rm inc}$ \cite{rakichGeneralTreatmentOptical2009, moskaletsScatteringMatrixApproach2011,liAdiabaticThermalRadiation2019} and producing the outgoing field $\vec{e}_{\rm out}(\bm{q}) = \hat{S}(\bm{q})\vec{e}_{\rm inc}$, \hl{which can be defined in a similar orthonormal basis as the ones for the incoming field mentioned above}.
We further assume that one has access to all the outgoing light, so the full scattering matrix is unitary $\hat{S}^\dagger(\bm{q}) \hat{S}(\bm{q}) = I$.

Under these 
%two 
conditions, we show in SM that the optical work $dW$ produced upon an infinitesimal change of parameters, $\bm{q} \rightarrow \bm{q}+d\bm{q}$, can be expressed in terms of the corresponding variation of the \hl{outgoing field $\vec{e}_{\rm out}(\bm{q})$}:
\begin{equation}
    \bm{F} = \frac{dW}{d\bm{q}} = - \frac{i}{\omega} \hl{\vec{e}^{\, \dagger}_{\rm out}(\bm{q}) \frac{d\vec{e}_{\rm out}(\bm{q})}{d\bm{q}}} \,.
    \label{eq:force}
\end{equation}
where $\bm{F}$ is the generalized force conjugate to the parameters $\bm{q}$. The physical interpretation of $\bm{F}$ depends on the nature of the drive.
If $\bm{q}$ represents the coordinates of a mechanically driven particle, then $\bm{F}$ is the optical force~\cite{rakichGeneralTreatmentOptical2009, wangResponseTheoryOptical2011}, and $dW$ represents the energy transferred from light to the kinetic motion of the particle after a small displacement. 
\hl{Magneto-electric coupling in moving matter~\cite{huidobroFresnelDragSpace2019} can be neglected in this adiabatic limit.}
In this case,  Eq.~(\ref{eq:force}) \hl{maps} to the familiar form of a generalized Wigner-Smith operator~\cite{ambichlFocusingDisorderedMedia2017, horodynskiOptimalWaveFields2020, hupflOptimalCoolingMultiple2023,horodynskiTractorBeamsOptimal2023} (originally introduced in mesoscopic physics~\cite{brouwerQuantumMechanicalTimeDelay1997a,DistributionQuantumMechanical1999}).
On the other hand, if $\bm{q}$ represents a set of electrically controlled refractive indices, then $\bm{F}$ stands for the intensities of the electric field at those points~\cite{horodynskiOptimalWaveFields2020}, and $dW$ is the energy transferred from the light to the material during the index change. This energy exchange is ultimately felt by the external drive. For example, for an increasing refractive index, the optical work assists the modulation and reduces the energy that must be supplied by an external battery by $dW$. 

\hl{The right-hand side of Eq.~(\ref{eq:force}) is proportional to the phase shift of the outgoing field in each channel after the perturbation, averaged by the respective outgoing intensity. This  intensity-averaged phase shift is the intensity-averaged Doppler frequency shift integrated over time, hence corresponding to the optical work (see SM for the proof). It has the same form as the Berry phase of a wavefunction, with the difference that in our scattering problem, the phase shift in each outgoing channel has no global phase ambiguity, and is completely determined by the incoming field and the scattering matrix.}

\begin{figure*}[ht]
    \centering
    \includegraphics[width=1\linewidth]{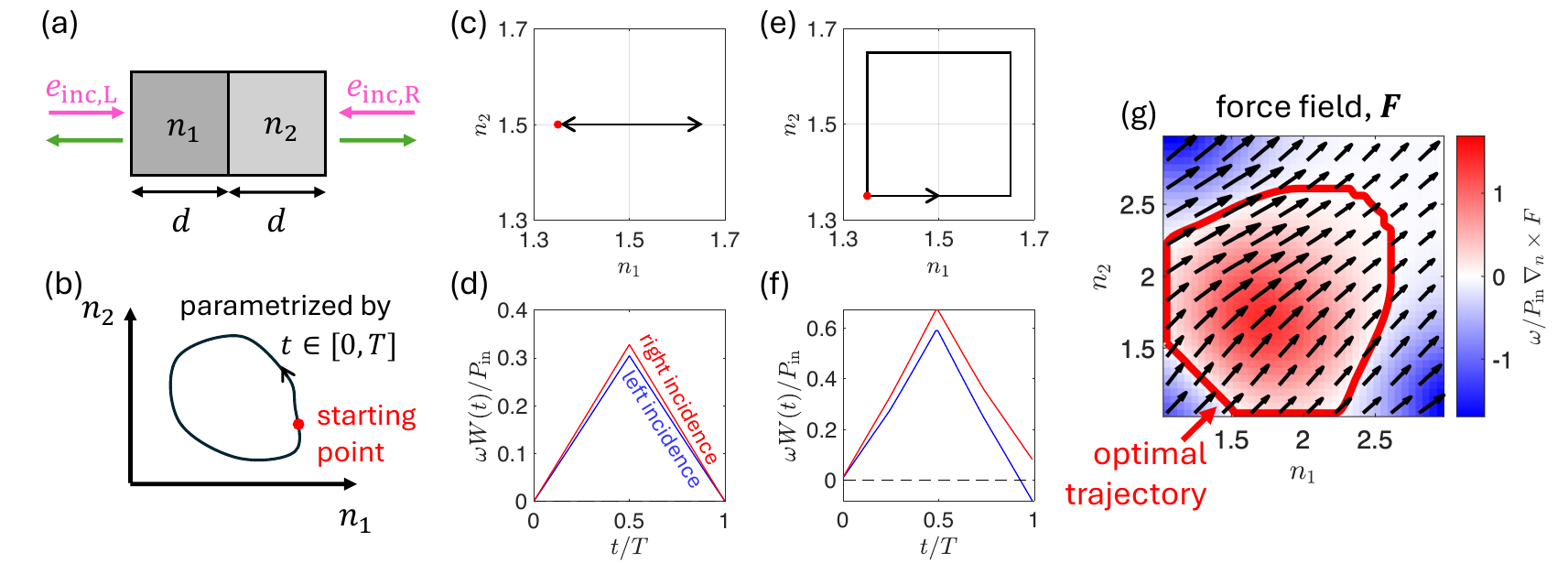}
    \caption{(a) Light scattering from a two-layer slab with tunable refractive indices $n_1$ and $n_2$ and equal layer thicknesses $d$. (b) The indices are varied adiabatically along a loop in parameter space $(n_1,n_2)$, parametrized by $t\in[0,T]$; the red dot marks the starting point. (c) A 1D loop yielding (d) zero net optical work after one period for both left and right incidence. (e) A 2D loop
    %, where both $n_1$ and $n_2$ vary significantly, 
    yielding (f) nonzero net optical work after one period. Left and right incidence produce equal and opposite works. (g) Generalized force field $\bm{F}$ (Berry connection) in parameter space, together with its curl (Berry curvature or the optical work density), and the optimal trajectory that maximizes the optical work per modulation cycle.}
    \label{fig:2}
\end{figure*}

%We show in Appendix this equation holds, not only for mechanically driven systems, but also for any systems whose permittivity is modulated adiabatically in time, including in particular the electrically modulated systems~[refs], which are of much current interest. 
%The operator $-i S^\dagger(\varepsilon)d\hat{S}(\bm{q})/d\varepsilon$ is a  general Wigner-Smith operator [refs]. Here, $d / d\varepsilon$ is a derivative over the permittivity's spatial distribution, and thus includes the previous instances of the general Wigner Smith operator (obtained from, e.g., particle translation, and permittivity increase at one point) as subsets.  

%The right-hand side of Eq.~(1) can be physically understood as follows: In unitary scattering, it is proportional to $\sum_{i=1}^N A_i^2 d\phi_i$, where $A_i$ and $\phi_i$ are the amplitude and phase of $e_{\textrm{out}, i}$. The amplitude term $A_i^2$ is proportional to the number of photons coming out of channel $i$ per second, the phase shift $d\phi_i$ is proportional to the Doppler frequency shift $\Delta \omega_i$ experienced by each photon coming out of that channel. Their product, summed over all channels, is the total energy lost by the photons, and hence the total work done to the external drive. 

For an extended parameter modulation along a trajectory $C$ in the $\bm{q}$-space, the total optical work becomes:
\begin{align}
    W&=\int\limits_C \bm{F}\cdot d\bm{q}  =
    \hl{\frac{1}{\omega}\,
    \int\limits_C -i
    \vec{e}^{\, \dagger}_{\rm out}(\bm{q})\,
    \frac{d\vec{e}_{\rm out}(\bm{q})}{d\bm{q}}\!\cdot d\bm{q}} \nonumber \\
    &= \frac{1}{\omega}\,\vec{e}^{\, \dagger}_{\rm inc}\!
    \left[
    \int\limits_C -i
    \hat{S}^\dagger(\bm{q})\,
    \frac{d \hat{S}(\bm{q})}{d\bm{q}}\!\cdot d\bm{q}
    \right]\!
    \vec{e}_{\rm inc}\,, \label{eq:work_loop_S}
\end{align}
\hl{where in the last step we replace the outgoing field with $\vec{e}_{\rm out}(\bm{q}) = \hat{S}(\bm{q})\vec{e}_{\rm inc}$.}
When the trajectory $C$ is closed, the contour integral in Eq.~\eqref{eq:work_loop_S} can be transformed into a surface integral using the generalized Stokes' theorem. In particular, for two time-varying parameters, $\bm{q}=(q_1,q_2)$, it takes the form
\begin{align}
    W &= \iint\limits_A \bm{\nabla}_{\bm{q}} \times \bm{F}\, dA \nonumber \\ 
    &= \frac{2}{\omega} \iint\limits_A \hl{\textrm{Im}\!\left[
    \frac{\partial \vec{e}_{\rm out}^{\, \dagger}(\bm{q})}{\partial q_1}\frac{\partial \vec{e}_{\rm out}(\bm{q})}{\partial q_2}
    \right]\!} dA,
\label{eq:work_surface}
\end{align}
where $A$ is the oriented area enclosed by the loop $C = \partial A$ [see Fig.~\ref{fig:1}(b)], and $dA = dq_1 dq_2$. 
Equations~(\ref{eq:work_loop_S}) and (\ref{eq:work_surface}) have mathematical forms similar to adiabatic charge pumping~\cite{citro_thouless_2023,brouwerScatteringApproachParametric1998,switkesAdiabaticQuantumElectron1999a} or the geometric amplitude (the imaginary Berry phase in non-Hermitian systems)~\cite{garrisonComplexGeometricalPhases1988, Bliokh1999JPA, Singhal2023PRR, Ozawa2025PRR, Lane2025}, where the integrands of the contour and surface integrals (i.e., the generalized force and its curl) play the role of the Berry connection and curvature, respectively. However, in contrast to adiabatic charge pumping, we deal with {\it coherent} scattering; and in contrast to the geometric amplitude, the system under consideration is {\it Hermitian} for any fixed value of parameters $\bm{q}$. 
\hl{
Particular instances of Eqs.~(\ref{eq:work_loop_S}) and (\ref{eq:work_surface}), for a two-channel time-varying waveplate and for a particle transported along a closed loop, were noted in Ref.~\cite{bliokhDynamicGeometricShifts2025}\ \footnote{\hl{In the relevant discussion of Ref.~\cite{bliokhDynamicGeometricShifts2025}, the present work, then in preparation,
was cited under the working title ``Maximizing
energy transfer in time-varying optical systems.''}}, in the context of a dynamic-geometric decomposition of the generalized Wigner-Smith operator for infinitesimal parameter changes. }
Equation~(\ref{eq:work_surface}) shows that the optical work accumulated during one modulation cycle is determined by the Berry curvature and the area $A$. Consequently, the optical work is independent of the modulation speed and depends only on the geometry of the trajectory. Furthermore, this work is {\it reversible} by reversing the direction of the parameter variation along the contour $C$ \cite{Bliokh1999JPA}. 
Equation~(\ref{eq:work_surface}) can be further generalized to $N$  parameters using 2-forms~\cite{spivakCalculusManifoldsModern2018}.

Equations~(\ref{eq:work_loop_S}--\ref{eq:work_surface}) constitute the central results of this work and have several important consequences. 
First, a non-zero optical work per cycle requires the independent variation of at least two parameters, such that the modulation loop encircles a non-zero area $A$ in parameter space [see Fig.~\ref{fig:1}(b)]
%Moreover, these variations must  [see Fig.~\ref{fig:1}(b)].
and thereby breaks the time-reversal symmetry of the drive~\cite{rakichGeneralTreatmentOptical2009, moskaletsScatteringMatrixApproach2011,liAdiabaticThermalRadiation2019}. 
%The \hl{effect} of time-reversal symmetry breaking by an adiabatic periodic drive can be characterized by the amount of energy transfer in Eqs.~(\ref{eq:work_loop}) and (\ref{eq:work_surface}).
This requirement applies, however, only to parameters tracing out a contractible loop in parameter space: a periodic angular coordinate that winds from 0 to 2$\pi$ can transfer net energy in each period even when varied alone.
%Note that this applies only to cyclically varying parameters that do not ``jump'' in their parameter space. For example, an angular coordinate that jumps from $2\pi$ to 0 can induce net energy transfer after a period. 
Likewise, non-adiabatic ``jumps'' of a single parameter in time can also produce optical net work \cite{Bliokh2002}.  

%Second, from the unitary character of the scattering matrix, it follows that $\textrm{Tr}\left[\oint_C \hat{S}^\dagger(\bm{q})\, d \hat{S}(\bm{q})\right] = 0$~\cite{guoSingularTopologyScattering2023,liuSpectralPhaseSingularity2023}. In adiabatic charge pumping, this property is directly related to the total charge conservation \cite{moskaletsScatteringMatrixApproach2011}. In our system, however, the trace operator represents an incoherent sum of optical work over all incoming channels at one frequency. This means that although individual incident modes may contribute either positive or negative optical work, an incoherent illumination with each incoming channel equally and incoherently occupied (i.e., the maximally mixed state with minimal purity) produces zero net optical work. 

\hl{Second, Eq.~(\ref{eq:work_loop_S}) has a surprising consequence when applied to incoherent illumination: Consider the extreme case of light coming in incoherently and with equal power from every channel (known as the maximally mixed state). Its optical work, according to Eq.~(\ref{eq:work_loop_S}), is the direct sum of the contributions from all channels, i.e., $W \propto \operatorname{Tr}\!\left[\oint_C \hat{S}^\dagger(\bm{q})\,d\hat{S}(\bm{q})\right]$. This integral is \textit{zero} for any unitary scattering matrix under contractible loops~\cite{guoSingularTopologyScattering2023,liuSpectralPhaseSingularity2023}. In adiabatic charge pumping, this is related to total charge conservation~\cite{moskaletsScatteringMatrixApproach2011}. In our case, it means an incoherent illumination with equal and incoherent incoming power in each channel  produces no net optical work after a contractible modulation cycle.}

%Second, the unitary character of the scattering matrix implies that $\operatorname{Tr}\!\left[\oint_C \hat{S}^\dagger(\bm{q})\,d\hat{S}(\bm{q})\right]=0$ \cite{guoSingularTopologyScattering2023,liuSpectralPhaseSingularity2023}. In adiabatic charge pumping, this property is directly related to total charge conservation \cite{moskaletsScatteringMatrixApproach2011}. \hl{To interpret it in our setting, we describe a partially coherent incident field by its normalized coherence matrix $\rho=\langle\vec{e}_{\rm inc}\vec{e}_{\rm inc}^{\,\dagger}\rangle/P_{\rm in}$, with $\operatorname{Tr}\rho=1$, and characterize its mixedness by the purity $\mathcal{P}=\operatorname{Tr}(\rho^2)$. For $M$ incoming channels, $1/M\leq\mathcal{P}\leq1$, with the minimum $\mathcal{P}=1/M$ attained by the maximally mixed state $\rho=I/M$, corresponding to equal and mutually incoherent population of all channels. The trace identity above, together with Eq.~(\ref{eq:work_loop_S}), shows that this minimum-purity illumination produces exactly zero net optical work over a modulation cycle.}

%This perfect cancellation among all incoming modes holds for any adiabatic and periodic modulation, regardless of the number of modulated parameters or the trajectory of the modulation.

Third, Eqs.~\eqref{eq:work_loop_S} and \eqref{eq:work_surface} allow one to optimize the incident field and modulation trajectory in parameter space to maximize (or minimize) the adiabatic optical work. For example, for 
a fixed trajectory $C$, Eq.~(\ref{eq:work_loop_S}) shows that the optical work is maximized when $\vec{e}_{\rm inc}$ is the eigenvector of the operator inside the square brackets with the maximum corresponding eigenvalue. This optimization procedure is analogous to optimizing optical force via the generalized Wigner-Smith operator~\cite{ambichlFocusingDisorderedMedia2017, horodynskiOptimalWaveFields2020, hupflOptimalCoolingMultiple2023,horodynskiTractorBeamsOptimal2023}, but now applied to the optical work along a trajectory.
\hl{This procedure applies also to partially incoherent incidence, and we show in SM that the more incoherent the incident field is (i.e., the less concentrated the spectrum of its coherence matrix is in terms of majorization order~\cite{guoTransportMeasurementsMajorization2025}), the less optical work one can expect to achieve through wavefront shaping, with the maximally mixed input above giving always zero optical work.}
\hl{On the other hand, in terms of optimizing the modulation trajectory, we see that} the integrand of Eq.~(\ref{eq:work_surface}) is proportional to the {\it density} of optical work in the parameter space. Hence, the optimal trajectory maximizing the work should encircle the maximum area of positive work density. 
%In particular, if one could identify in the parameter space a \textit{patch} of positive (or negative) contributions, then the loop that encloses this patch should give the maximal optical work per cycle.

\begin{figure*}[ht]
    \centering
    \includegraphics[width=1\linewidth]{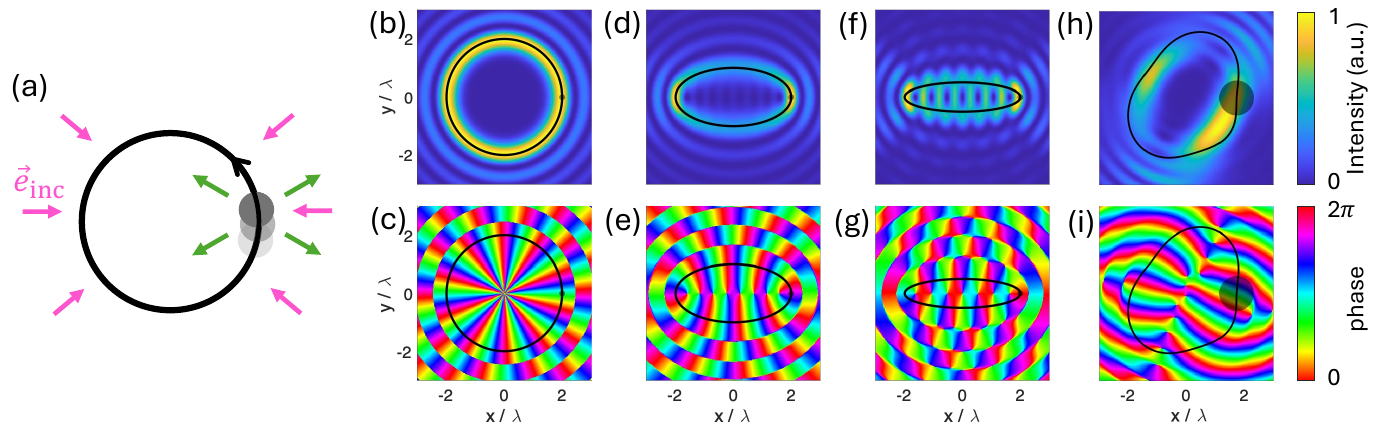}
    \caption{(a) Light scattering from a particle moving along a prescribed closed trajectory in real space. 
    (b--i) Intensity and phase distributions of the selected components of optimal incident electric field that maximizes optical work along the loop (see explanations in the text).  
    %for a particle of radius $0.1\lambda$ in (b,c) a circular trajectory, (d,e) an elliptical trajectory, (f,g) a strongly compressed elliptical trajectory, and (h,i) a particle of radius $0.6\lambda$ in an asymmetric trajectory. 
    The optimal fields have high intensities and phase gradients along the particle trajectory, which maximize the nonconservative radiation-pressure force \cite{toftulRadiationForcesTorques2026, berryPhysicalCurlForces2013, sukhovNonconservativeOpticalForces2017}.}
    \label{fig:3}
\end{figure*}

{\it Examples.---} We now illustrate the main features of our general approach with two examples. First, we consider 1D scattering from a two-layer dielectric slab [see Fig.~\ref{fig:2}(a)]. The two layers have refractive indices $n_1$ and $n_2$, % respectively, 
and each has thickness $d=\lambda/(2\pi)$ where $\lambda$ is the vacuum wavelength of the incident light. The field $\vec{e}_{\rm inc}$ is normally incident from either left or right, defining the two scattering channels of the problem: $\vec{e}_{\rm inc} = \left( \begin{matrix} e_{\rm inc, L} \\ e_{\rm inc, R} \end{matrix}\right)$. We vary $(n_1,n_2)$ along a counterclockwise loop in parameter space [see Fig.~\ref{fig:2}(b)], parametrized by $t\in[0,T]$, with $T$ being the period of the modulation. By $W(t)$ we denote the optical work accumulated from the starting point $t_0=0$ to the point $t$. The scattering matrix $\hat{S}(n_1,n_2)$ is computed using the Fresnel coefficients for each interface (see SM), and the optical work $W(t)$ is then obtained from Eq.~(\ref{eq:work_loop_S}). We normalize $W(t)$ by the incoming energy per radian, $P_{\rm in}/\omega$, where $P_{\rm in}=\vec{e}^{\, \dagger}_{\rm inc} \vec{e}_{\rm inc}$ is the incoming power, so that the relevant dimensionless quantity is $\omega W(t)/P_{\rm in}$. We first consider a 1D loop shown in Fig.~\ref{fig:2}(c), \hl{where $n_1$ changes from 1.35 to 1.65 and then back to 1.35, while $n_2$ stays at 1.5. We find the total work $W(t)$ vanishes at the end of the loop (up to machine precision)} for incidence from either side, $\vec{e}_{\rm inc} \propto \left( \begin{matrix} 1 \\ 0 \end{matrix}\right)$ or $\left( \begin{matrix} 0 \\ 1 \end{matrix}\right)$. \hl{The evolution of $W(t)$ during the loop is shown in Fig.~\ref{fig:2}(d).}
%Figure~\ref{fig:2}(d) shows that the net optical work vanishes \hl{(up to machine precision in our numerical result)} after one period for incidence from either side, $\vec{e}_{\rm inc} \propto \left( \begin{matrix} 1 \\ 0 \end{matrix}\right)$ or $\left( \begin{matrix} 0 \\ 1 \end{matrix}\right)$. 
In contrast to the 1D loop, a 2D loop shown in Fig.~\ref{fig:2}(e) produces a nonzero net optical work after one period [see Fig.~\ref{fig:2}(f)]. Furthermore, Fig.~\ref{fig:2}(f) shows that the incident light modes coming from the left and from the right produce equal and opposite optical works, so their incoherent sum vanishes \hl{(up to machine precision in our numerical results)}, in agreement with our prediction.

Next, Fig.~\ref{fig:2}(g) displays the generalized force field $\bm{F}$ over the parameter space $(n_1,n_2)$, calculated via Eq.~(\ref{eq:force}) for the right-incident field. 
The components of this field represent the optical works $dW_1$ and $dW_2$ performed by the light field on the medium under infinitesimal changes of the two refractive indices, $dn_1$ and $dn_2$. All force vectors point toward increasing $n_1$ or $n_2$, because light always performs positive work when the refractive index increases. It is not clear from the force field, however, how one should periodically modulate the refractive indices to maximize the work. Remarkably, by plotting the {\it curl} of the force field, i.e., the integrand 
in Eq.~(\ref{eq:work_surface}), one can clearly identify regions of positive [the red area in Fig.~\ref{fig:2}(g)] and negative [blue areas] optical work density. \hl{Thus, an illustrative choice of modulation trajectory is the loop indicated by the red curve in Fig.~\ref{fig:2}(g), which encloses the positive-density region while excluding all the negative-density regions. Among all simple loops (i.e., loops that do not cross themselves) contained within the plotted region of Fig.~\ref{fig:2}(g), this loop maximizes the optical work per cycle.} 

%Thus, \hl{in the plotted region of Fig.~\ref{fig:2}(g) and assuming the trajectory being a simple loop (a loop that does not cross itself)}, the optimal trajectory that maximizes the optical work \hl{after a complete loop} is the one that encloses the positive-density region while excluding all the negative-density regions, as indicated by the red curve in Fig.~\ref{fig:2}(g). 

As the second example, we examine 3D light scattering from a dielectric spherical particle [see Fig.~\ref{fig:3}(a)]. The center of the particle, described by position $\bm{r}$, moves along a prescribed loop $C$ in real space. 
For each position of the particle, $\bm{r}\in C$, we compute the scattering matrix $\hat{S}(\bm{r})$ using the $T$-matrix formalism implemented in the Optical Tweezer Toolbox~\cite{nieminenOpticalTweezersComputational2007}. The associated scattering channels are the vector spherical waves defined with respect to the center of the loop. 
We obtain the desired incident field $\vec{e}_{\rm inc}$ that maximizes the optical work along $C$ from the eigenproblem in Eq.~(\ref{eq:work_loop_S}). For a sphere with refractive index $n=3$ and radius $0.1\lambda$, moving along a circle of radius $2\lambda$ in the $(x,y)$-plane [see Figs.~\ref{fig:3}(b,c)], the optimal incident field is $z$-polarized and corresponds to a vortex wave having maximum intensity on the loop and propagating azimuthally along the loop \cite{Ceperley1992AJP, Allen_book, Garces-Chavez2003PRL, Wang2025N}. By changing the sign of this vortex (i.e., the azimuthal-propagation direction), one can reverse the sign of the optical work. 
Notably, the nonzero optical work for a small particle transported along a closed contour $C$ reflects the non-conservative nature of the radiation-pressure optical force directed along the phase gradient \cite{toftulRadiationForcesTorques2026, berryPhysicalCurlForces2013, sukhovNonconservativeOpticalForces2017}, and has a geometric origin \cite{bliokhDynamicGeometricShifts2025}.

Next, we change the circular path into an ellipse. The optimal incident field follows this transformation, forming an elliptical intensity maximum and phase gradient along the trajectory [see Figs.~\ref{fig:3}(d,e)]. 
When we compress the ellipse further, the two sides meet and produce interference fringes [see Figs.~\ref{fig:3}(f,g)]. 
\hl{This optimal field produces an optical work 2.3 times larger than that from the best vortex beam (see SM).}
To demonstrate that our theory applies to arbitrary trajectories and particle sizes, we consider an asymmetric trajectory for a larger sphere of radius $0.6\lambda$ [see Figs.~\ref{fig:3}(h,i)]. The resulting optimal field becomes predominantly $x$- and $y$-polarized; the phase of the $x$-component is shown in Fig.~\ref{fig:3}(i). The field is concentrated on two sides of the trajectory, with large phase gradients that maximize the optical work per cycle.%, similar to the previous ones in Fig.~\ref{fig:3}(b-g).  

%Figure~\ref{fig:3}(j) compares the optical work $W$ done by the optimal incident field after a cycle with that from non-optimized plane-wave incidences for the loop in Figs.~\ref{fig:3}(h,i). The normalized work $\omega W/P_{\rm in}$ is shown as the red bar for the optimal field and as blue bars for the plane waves. Each blue bar corresponds to a specific incident angle and polarization. We use 80 plane waves to discretize all incident directions, spanning 5 polar angles, 8 azimuthal angles, and 2 polarizations per direction. The optimal field produces more than two orders of magnitude larger optical work than any individual plane wave. Moreover, the plane-wave contributions include both positive and negative values and numerically sum to zero, consistent with our prediction that incoherent illumination produces no net work after one modulation cycle.

{\it Conclusions.---}To summarize, we have established a general geometric theory of optical work in coherent wave scattering and illustrated its remarkable features using representative examples from time-varying photonics and optomechanics. Specifically, we found that the net optical work over a closed loop in parameter space is associated with contour or surface integrals of the corresponding Berry connection or curvature; therefore, it requires the modulation of at least two independent parameters.   
%In adiabatic charge pumping, this time-reversal symmetry breaking leads to charge pumping~\cite{brouwerScatteringApproachParametric1998}; in our case, it leads to the pumping of optical work. 
%In both cases, the scattering operator in Eq.~(\ref{eq:work_surface}) characterizes the degree of time-reversal symmetry breaking.  
%This operator generalizes the concept of nonconservative optical forces in optomechanics~\cite{rakichGeneralTreatmentOptical2009, berryPhysicalCurlForces2013,sukhovNonconservativeOpticalForces2017} to general time-varying optical systems involving, for example, electrical, thermal, or optical modulations. 
Notably, a field that \hl{equally and incoherently populates all incoming channels} produces zero net optical work after a contractible modulation cycle. For optomechanics,  this implies that, within our assumptions of linear, lossless scattering and adiabatic motion, the cycle-averaged optical force from thermal light – and, by extension, from vacuum fluctuations, which populate all scattering channels at each frequency with equal weight \hl{(see SM for details)} – is conservative. 
%This is consistent with non-conservative quantum-friction forces \cite{Pendry_1997,PhysRevLett.105.113601}, which are velocity-dependent and therefore vanish in the adiabatic limit considered here.
\hl{Our theory also provides a theoretical basis for the previously observed adverse effect of spatial incoherence on the non-conservative force~\cite{sukhovNonconservativeOpticalForces2017}, and generalizes the existing numerical results on small particles~\cite{wangEffectSpatialCoherence2007, zenkovaOpticalCurrentsVector2011, aunonOpticalForcesSmall2012, sukhovNonconservativeOpticalForces2017} to arbitrary linear, lossless electromagnetic scattering under incoherent and equal-channel excitation.}
 We also demonstrated, for coherent incident light, how the optical work can be maximized by optimizing the incident wavefront and the modulation trajectory in parameter space. 
 %\hl{The benefit of wavefront shaping also applies to partially incoherent light field: the more coherence in the incident field, the more optical work one can generate with wavefront shaping.}
This extends wavefront-shaping~\cite{rotterLightFieldsComplex2017, caoShapingPropagationLight2022} and inverse-design~\cite{lalau-keralyAdjointShapeOptimization2013, moleskyInverseDesignNanophotonics2018} concepts from static to time-varying optical systems.  

%It would also be interesting to explore the topological aspects of multi-parameter modulation, such as geometric phases~\cite{bliokhDynamicGeometricShifts2025}, in analogy with past studies of adiabatic quantum pumping~\cite{avronGeometryStatisticsAsymptotics2000}, and to examine possible topological signatures in optical work near the scattering singularities in the aforementioned non-Hermitian systems~\cite{guoSingularTopologyScattering2023,liuSpectralPhaseSingularity2023}. For applications in time-varying photonic media, a natural question is how to further enhance the magnitude of the optical work, for example by exploiting resonant effects or by increasing the size of the structure and the number of its tunable parameters. In the limit of infinitely many tunable parameters, we expect our theory to include the existing traveling-wave-like modulation~\cite{galiffiBroadbandNonreciprocalAmplification2019, pendryGainTimedependentMedia2021, galiffiArchimedesScrewLight2022} and its variations ~\cite{mostafaBroadbandAmplificationLight2025}, in which exponentially large energy transfer between light and the external drive has been observed. 
The adiabatic nature of our optical-work pump is particularly attractive for time-varying photonics, as it provides an experimentally feasible and reversible method to control the energy exchange between light and the external drive via {\it slow} modulation. This contrasts with the existing fast driving schemes which are experimentally challenging, especially at optical frequencies~\cite{zhouBroadbandFrequencyTranslation2020, miyamaruUltrafastFrequencyShiftDynamics2021, moussaObservationTemporalReflection2023, lyubarovAmplifiedEmissionLasing2022, michaelPhotonicTimecrystallineBehaviour2024, galiffiOpticalCoherentPerfect2026}. 
\hl{Our optimal modulation loop could be extended to more complex loops that cross themselves, potentially with additional constraints on the total loop length.}  
In the optomechanical context, our approach can be used for efficient manipulation of particles~\cite{horodynskiOptimalWaveFields2020, toftulRadiationForcesTorques2026} or light-driven nano-motors~\cite{xu_light-driven_2017}. Traditionally-used vortex or circularly polarized beams are mostly optimized for circular geometries and small particles; our formalism allows one to numerically identify the optimal illumination for arbitrary trajectories and particles. \hl{This numerical procedure extends to the case where one does not have a complete control of all incident channels, such as typical optical tweezers with NA = 1.1--1.3~\cite{butaitePhotonefficientOpticalTweezers2024}.}
%In this case, one should replace the full scattering matrix in Eq.~(\ref{eq:work_loop_S}) that map all incoming channels to all outgoing channels with a sub scattering matrix that map a set of experimentally available incoming channels to all outgoing channels to derive the optimal incident field.} 
%A natural next step is to relax the fixed-trajectory assumption and self-consistently incorporate the optical force's back-action on the particle's motion. 

Our theoretical and numerical results can be experimentally tested in a number of platforms, including optical tweezers~\cite{drobyOpticalTweezersLight2025, toftulRadiationForcesTorques2026}, water waves~\cite{orazbayevWavemomentumShapingMoving2024, Wang2025N}, and acoustic waves~\cite{eggletonBrillouinIntegratedPhotonics2019, toftulRadiationForcesTorques2026}, as well as electrically-modulated metasurfaces~\cite{zhangSpacetimecodingDigitalMetasurfaces2018, guoNonreciprocalMetasurfaceSpace2019} and coupled photonic resonators~\cite{liAdiabaticThermalRadiation2019}.
Further development of the proposed approach might include generalizations to nonadiabatic and nonperiodic drives, non-Hermitian and nonlinear optical media, and explore potential quantization behavior akin to that arising in Thouless pumping~\cite{thouless_quantization_1983, citro_thouless_2023}.

\hl{{\it Data Availability Statement.---}
The codes used to generate figures in the manuscript and SM are available from https://doi.org/10.5281/zenodo.23809806.}

\begin{acknowledgments}
Z.K. was supported by the Austrian Science Fund (FWF) [10.55776/COE1]. 
K.Y.B. was supported by Grant PID2025-169574NB-C21 funded by MICIU/AEI/10.13039/501100011033 and co-funded by the European Union through the project HORIZON-MSCA-2022-COFUND-01-SmartBRAIN3-101126600.
S.R. was supported by the Austrian Science Fund (FWF) [10.55776/PIN7240924].
We thank David Globosits, Jakob Hüpfl, Tsampikos Kottos, Luca Neubacher, and Arunn Suntharalingam for helpful discussion.
\end{acknowledgments}

\bibliography{ref.bib}

\clearpage
\section*{Supplemental Material}

\setcounter{figure}{0}
\renewcommand{\thefigure}{S\arabic{figure}}
\renewcommand{\theHfigure}{supp.\arabic{figure}}

\section{Energy transfer in the adiabatic limit}
\label{sec:energy-transfer}

We show in this section that the optical work in the near field can be expressed in terms of far-field quantities as in Eq.~(1) of the main text. Our analysis largely follows Ref.~\cite{rakichGeneralTreatmentOptical2009}.

We consider light coming into a complex scattering system [e.g., Fig.~1(a) in the main text] through $N$ orthonormal ports at frequency $\omega$ with a total power $P_{\rm in}$, and leaving these ports with frequency $\omega_{{\rm out}, i}(t)$ and power $P_{{\rm out}, i}(t)$ for port $i=1,2,\dots,N$. These output quantities are functions of time because the scatterer changes in time. During this time, two other things will happen: the energy stored in the system $U_{\rm in}(t)$ may increase or decrease; light might also do some work $W(t)$ on the time-varying medium. These are the key power quantities whose inter-relation we will elucidate in this section.

Our starting point is energy and photon number conservation. Energy conservation says the input power equals the output power, plus the rate of energy increase inside the system, plus the rate of external work:
\begin{equation}
    P_{\rm in} = \sum_{i=1}^N P_{{\rm out}, i}(t) + \frac{dU_{\rm in}(t)}{dt} + \frac{dW(t)}{dt}\,.
    \label{eq:s1}
\end{equation}
Photon number conservation says the number of photons coming in per unit time equals the total number of photons coming out per unit time, plus the increase in the number of photons inside the system per unit time. We can count the number of photons by dividing the corresponding power quantities by their instantaneous frequencies, so we have:
\begin{equation}
    \frac{P_{\rm in}}{\omega} = \sum_{i=1}^N \frac{P_{{\rm out}, i}(t)}{\omega_{{\rm out}, i}(t)} + \frac{d}{dt}\frac{U_{\rm in}(t)}{\omega_{\rm in}(t)},
    \label{eq:s2}
\end{equation}
where $\omega_{\rm in}(t)$ is the instantaneous frequency of the light stored inside the system.
Both equations above can be derived from Maxwell's equations. Equation~\eqref{eq:s1} can be derived from the well-known Poynting theorem~\cite{jacksonClassicalElectrodynamics2012}. Equation~\eqref{eq:s2} can be derived from action conservation~\cite{globositsPseudounitaryFloquetScattering2024}.

\begin{figure}[h!]
    \centering
    \includegraphics[width=\linewidth]{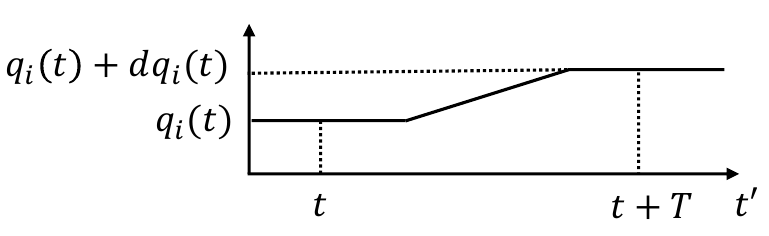}
    \caption{Adiabatic time evolution starting from a steady state with control parameters $\{q_i(t)\}_{i=1}^N$ and ending at a steady state with control parameters $\{q_i(t)+dq_i(t)\}_{i=1}^N$.}
    \label{fig:s1}
\end{figure}

What we are interested in is not the instantaneous response of the time-varying system but the overall energy transfer when it evolves from one steady state to another steady state. In particular, we consider an evolution of the system following the diagram in Fig.~\ref{fig:s1}, where the control parameters $\{q_i(t)\}_{i=1}^N$ first stay at one set of specific values (a steady state), then change slightly, then stay at the slightly changed values (another steady state). We ask how much optical work is done from one time $t$ in the first steady state to another time $t+T$ in the other steady state. To calculate this, we integrate Eqs.~\eqref{eq:s1} and \eqref{eq:s2} along this time period, which yields
\begin{align}
    P_{\rm in} T &= \sum_{i=1}^N\int_t^{t+T} P_{{\rm out}, i}(t')\,dt' + \Delta U_{\rm in} + \Delta W\,, \label{eq:s3}\\
    \frac{P_{\rm in}}{\omega}T &= \sum_{i=1}^N\int_t^{t+T} \frac{P_{{\rm out}, i}(t')}{\omega_{{\rm out}, i}(t')}\,dt' + \frac{\Delta U_{\rm in}}{\omega}\,. \label{eq:s4}
\end{align}
From these two equations we can solve $\Delta W$ and $\Delta U_{\rm in}$ in terms of the other quantities:
\begin{align}
    \Delta W &= \sum_{i=1}^N\int_t^{t+T} \frac{P_{{\rm out}, i}(t')}{\omega_{{\rm out}, i}(t')}
    \left[\omega-\omega_{{\rm out}, i}(t')\right]dt'\,, \label{eq:s5}\\
    \Delta U_{\rm in} &= \left[\frac{P_{\rm in}}{\omega}T -
    \sum_{i=1}^N\int_t^{t+T}\frac{P_{{\rm out}, i}(t')}{\omega_{{\rm out}, i}(t')}\,dt'\right]\omega\,. \label{eq:s6}
\end{align}
The first equation shows that the total work equals the Doppler shift in the outgoing light. The second equation shows that the increase in the stored energy comes from the photons that choose to stay in the system. In other words, light upon scattering can lose its energy in two ways: either a frequency cut or an amplitude cut. Energy lost from the frequency cut will do work; energy lost from the amplitude cut will just be stored in the system.

We have now established that the optical work in the near field translates to Doppler shift in the far field. But to measure this work, as suggested by Eq.~\eqref{eq:s5}, we need to constantly track the outgoing power and frequency, which is inconvenient. It would be much better if we only needed to measure the system's responses at some steady states, say at $t$ and $t+T$, and deduce work $\Delta W$ from there.

This can indeed be done as long as we change the system slowly. In fact, in the adiabatic limit, we can replace $P_{{\rm out}, i}(t')/\omega_{{\rm out}, i}(t')$ with $P_{{\rm out}, i}(t)/\omega_{{\rm out}, i}(t)$ with no penalty~\cite{rakichGeneralTreatmentOptical2009}. This turns Eq.~\eqref{eq:s5} into
\begin{equation}
    \Delta W = \frac{1}{\omega}\sum_{i=1}^N P_{{\rm out}, i}(t)
    \int_t^{t+T}\left[\omega-\omega_{{\rm out}, i}(t')\right]dt'.
    \label{eq:s7}
\end{equation}
The integral in the equation above is the phase lag of the outgoing light resulting from the frequency shift. Defining $d\phi_i(t)/dt=\omega-\omega_{{\rm out}, i}$, we have
\begin{equation}
    \Delta W = \frac{1}{\omega}\sum_{i=1}^N P_{{\rm out}, i}(t)
    \left[\phi_i(t+T)-\phi_i(t)\right].
    \label{eq:s8}
\end{equation}
We can write this in a more compact form by collecting the complex-valued amplitudes of the outgoing fields at the $N$ channels and packing them into the $N\times1$ vectors: $\vec{e}_{\rm out}$ for the outgoing field at the steady state $t$, and $\vec{e}_{\rm out}+d\vec{e}_{\rm out}$ for the outgoing field at the steady state $t+T$. This yields
\begin{equation}
    \Delta W = -\frac{i}{\omega}\vec{e}^{\,\dagger}_{\rm out}d\vec{e}_{\rm out}\,,
    \label{eq:s9}
\end{equation}
which is the form used in Eq.~(1) of the main text.

\section{Optical work and changes in permittivity}
\label{sec:permittivity}

In the previous section, we showed that Doppler shift in the far field equals optical work in the near field. In this section, we express the latter in terms of how permittivity changes in time, which is the natural way to describe examples like moving particles and electrically tuned refractive indices.

Physically, permittivity measures how easily dipoles in a unit volume can be polarized. When permittivity increases, the dipoles become more polarized. The two ends of a dipole, the negative and positive charges, move in opposite directions, creating a current. This current is proportional to the local field strength and how fast the permittivity changes:
\begin{equation}
    \Jv(\rv,t) = [\partial_t\varepsilon(\rv,t)]\Ev(\rv,t)\,.
    \label{eq:s10}
\end{equation}
Alternatively, this expression can be rigorously derived from Maxwell's equations. Assuming a non-magnetic material and instantaneous electric response $D(\rv,t)=\varepsilon(\rv,t)E(\rv,t)$, the Amp\`ere--Maxwell law becomes
$\nabla\times H(\rv,t)=J_f(\rv,t)+[\partial_t\varepsilon(\rv,t)]E(\rv,t)+\varepsilon(\rv,t)\partial_tE(\rv,t)$,
showing $[\partial_t\varepsilon(\rv,t)]E(\rv,t)$ as an effective free current generated by the changing permittivity. We further assume the permittivity is real-valued, so there is no absorption in the material.

When we have a current, light does work on the current. This is the work $W(t)$ we have previously phenomenologically included in the energy conservation of Eq.~\eqref{eq:s1}. The rate of the work is proportional to the spatial overlap between the current density and the electric field:
\begin{equation}
    \frac{dW(t)}{dt} = \int \Ev(\rv,t)\cdot\Jv(\rv,t)\,dV,
    \label{eq:s11}
\end{equation}
where $V$ is the whole space. Plugging Eq.~\eqref{eq:s10} into Eq.~\eqref{eq:s11}, and integrating both sides from $t$ to $t+T$, we have
\begin{equation}
    \Delta W = \int dV\int_t^{t+T}dt'\,
    \Ev(\rv,t')\cdot\frac{d\varepsilon(\rv,t')}{dt'}\Ev(\rv,t')\,.
    \label{eq:s12}
\end{equation}
All this work transfers to the Doppler shift in the far field. In the adiabatic limit, according to Eq.~\eqref{eq:s9}, we have
\begin{equation}
    -\frac{i}{\omega}\vec{e}^{\,\dagger}_{\rm out}d\vec{e}_{\rm out}
    =
    \frac{1}{2}\int dV\int_t^{t+T}dt'\,
    |\Ev(\rv,\varepsilon(\rv,t'))|^2
    \frac{d\varepsilon(\rv,t')}{dt'},
    \label{eq:s13}
\end{equation}
where we have converted, on the right-hand side, $\Ev(\rv,t')$ to its complex-valued equivalent:
$\Ev(\rv,t')=\operatorname{Re}[\Ev(\rv,\varepsilon(\rv,t'))e^{-i\omega_{\rm in}(t')t'}]$.
Because the permittivity is real-valued and varies slowly, we can replace
$\Ev(\rv,t')\cdot\Ev(\rv,t')$ with its temporal average
$|\Ev(\rv,\varepsilon(\rv,t'))|^2/2$.
The resulting equation, Eq.~\eqref{eq:s13}, is a general equation that relates the permittivity changes in the near field to the Doppler shift in the far field.

\begin{figure*} [ht]
    \centering
    \includegraphics[width=1\linewidth]{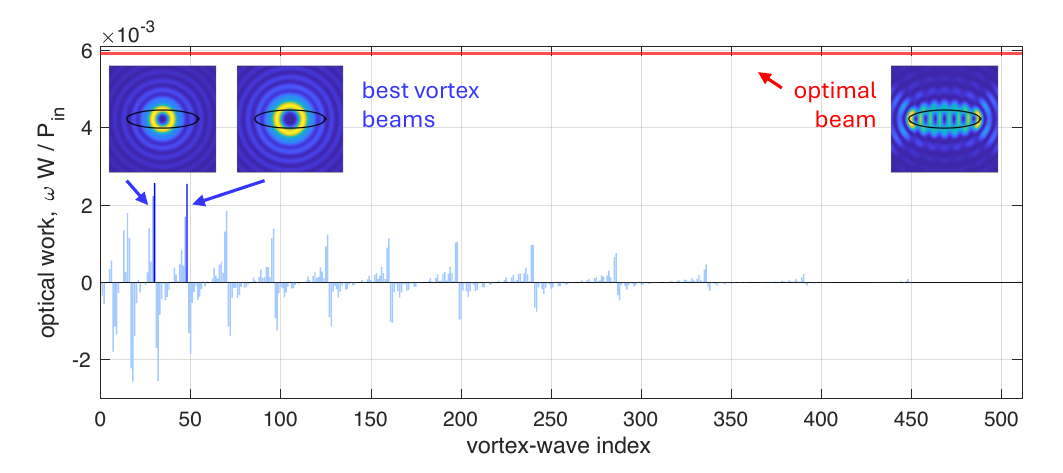}
    \caption{\hl{Optical work generated by the optimized beam and by different possible vortex beams.}} 
    \label{fig:5}
\end{figure*}

In some cases, we can simplify Eq.~\eqref{eq:s13} even further. Consider the case where the change in the field is not large, at least not large at the points where $\varepsilon(\rv,t')$ changes significantly. In this case, we can take the electric field outside of the integral and represent it by its value at time $t$. What is left in the integral is $d\varepsilon(\rv,t')/dt'$, which we can integrate from beginning to end:
\begin{equation}
    -\frac{i}{\omega}\vec{e}^{\,\dagger}_{\rm out}d\vec{e}_{\rm out}
    =
    \frac{1}{2}\int dV\,|\Ev(\rv,\varepsilon(\rv,t))|^2
    [\varepsilon(\rv,t+T)-\varepsilon(\rv,t)].
    \label{eq:s14}
\end{equation}
This result, reported previously in Ref.~\cite{horodynskiOptimalWaveFields2020}, connects steady state to steady state. What we reveal in our derivation is that the underlying assumption is that the electric field should not change too much at the points where the permittivity changes. This assumption breaks down when, for example, a Fabry--P\'erot cavity is moved to an on-resonance position. In this case, Eq.~\eqref{eq:s13}, which applies to any permittivity that changes adiabatically, should be used.

\section{Scattering matrix of the two-layer film}
\label{sec:two-layer}

The scattering matrix of the two-layer film in Fig.~2(a) of the main text is calculated in the following way. First, we calculate the transmission and reflection coefficients of each layer in vacuum:
\begin{equation}
t_i =
\frac{
\left(1-\eta_i^2\right)e^{ikn_i d_i}
}{
1-\eta_i^2e^{2ikn_i d_i}
}\,,\qquad
r_i =
\frac{
\eta_i\left(1-e^{2ikn_i d_i}\right)
}{
1-\eta_i^2e^{2ikn_i d_i}
}\,,
\label{eq:s15}
\end{equation}
where $i=1,2$ labels the two layers, $n_i$ is the refractive index of the $i$th layer, $d_i$ is its thickness, $\eta_i=(1-n_i)/(1+n_i)$, and $k$ is the vacuum wavevector.

Then we combine the single-layer coefficients above to form the reflection and transmission coefficients of the two-layer film~\cite{yehOpticalWavesLayered1988}. The reflection coefficient for waves coming to the two layers from the left is
\begin{equation}
r_L=r_1+\frac{t_1^2r_2}{1-r_1r_2}.
\label{eq:s16}
\end{equation}
The reflection coefficient for waves coming from the right is
\begin{equation}
r_R=r_2+\frac{t_2^2r_1}{1-r_1r_2}.
\label{eq:s17}
\end{equation}
The transmission coefficient (from either side) is
\begin{equation}
t=\frac{t_1t_2}{1-r_1r_2}.
\label{eq:s18}
\end{equation}
These coefficients make up the scattering matrix $\hat{S}$ of the two-layer film:
\begin{equation}
\hat{S} =
\begin{pmatrix}
r_L & t\\
t & r_R
\end{pmatrix}.
\label{eq:s19}
\end{equation}

\hl{
\subsection{Optimal beam vs. vortex beams}
This section compares the performance of the optimal beam in Fig.~3(f) of the main text with the traditionally used vortex beams. Under the same scattering configuration as Fig.~3(f) of the main text, we send in vortex beams of different orders, and calculate their resulting optical work after a modulation cycle. 

The result is summarized in Fig.~\ref{fig:5}. The x-axis runs through 510 vortex beams with radial number $l$ from 1 to 15, azimuthal number from $-l$ to $l$, and two possible polarizations. 
The best vortex beams are highlighted in dark blue bars (further increasing the range of $l$ does not yield better vortex beams). Their optical work is around $W=2.6\times10^{-3} P_{\rm in}/\omega$. The insets show the intensity profiles of their electric fields. In each profile, the upper and lower sides of the brightest ring align with the elliptical trajectory, therefore generate large work.
In contrast, the optimal beam presented in Fig.~3(f) of the main text, shown here as the red line, produces an optical work $W=5.9\times10^{-3} P_{\rm in}/\omega$, which is 2.3 times larger than that of the best vortex beams.}

\hl{
\section{Conservative character of the vacuum force in lossless systems}
\label{sec:vacuum-force}

In the main text, we state that the cycle-averaged force produced by vacuum fluctuations is conservative under the assumptions of linear, lossless scattering and adiabatic motion. Here we prove this and clarify how the monochromatic equal-channel cancellation extends to the broadband problem of vacuum fluctuation.

First, we write the vacuum force in terms of the frequency integral of the trace of the Wigner-Smith operator [see Eq.~(66) in Ref.~\cite{rachbauerHowFindOptimal2024a}]:
\begin{equation}
\bm{F}
=
-\frac{i\hbar}{4\pi}
\int_0^\infty
\operatorname{Tr}\left[
\hat{S}^\dagger(\mathbf{r},\omega)
\frac{d\hat{S}(\mathbf{r},\omega)}{d\mathbf{r}}
\right]d\omega
\end{equation}
where $\hat{S}(\mathbf{r},\omega)$ is the scattering matrix at frequency $\omega$ and with parameter $\mathbf{r}$ (say a particle's position). The optical work $W$ after a loop in the $\mathbf{r}$-space is
\begin{equation}
W
=
\oint \bm{F}\cdot d\mathbf{r}
=
-\frac{i\hbar}{4\pi}
\int_0^\infty d\omega
\oint
\operatorname{Tr}\left[
\hat{S}^\dagger(\mathbf{r},\omega)
\frac{d\hat{S}(\mathbf{r},\omega)}{d\mathbf{r}}
\right]\cdot d\mathbf{r}
\label{eq:vacuum}
\end{equation}
where we have changed the order of the integrals. Last, we invoke the theorem that for unitary scattering matrix the loop-integral above is zero for a contractible loop~\cite{guoSingularTopologyScattering2023,liuSpectralPhaseSingularity2023}:
\begin{equation}
\oint
\operatorname{Tr}\left[
\hat{S}^\dagger(\mathbf{r},\omega)
\frac{d\hat{S}(\mathbf{r},\omega)}{d\mathbf{r}}
\right]\cdot d\mathbf{r}
=0.
\label{eq:traceless}
\end{equation}
Therefore, the total work after a loop is also zero
\begin{equation}
W=\oint \bm{F}\cdot d\mathbf{r}=0.
\end{equation}
As this is true for any contractible loop, we conclude that the vacuum force is a conservative force in unitary scattering systems.

The above argument requires (i) a linear electromagnetic response, which may be dispersive; (ii) losslessness, so that the complete scattering matrix is unitary at each real frequency; (iii) adiabatic variation of the parameters; and (iv) sufficient regularity for exchanging the frequency and contour integrals in Eq.~(\ref{eq:vacuum}). 
}

\hl{
\section{Shaping partially incoherent light for maximal optical work}
\label{sec:incoherent}

The optical work formula in Eq.~(2) of the main text assumes a definite
coherent input. Here we generalize to partially incoherent illumination and, using the majorization technique~\cite{guoTransportMeasurementsMajorization2025}, show how coherence affects the maximum achievable work after wavefront shaping.

\textit{Average optical work.---}We describe the state of a partially incoherent incident field, before it reaches to a wavefront shaper, by a coherence matrix $\hat{\rho}_0$. This matrix is Hermitian, positive semidefinite, and satisfies $\operatorname{Tr}(\hat{\rho}_0)=P_{\rm in}$. It is rank-1 if the incident field is fully coherent, and equals to $(P_{\rm in}/N)\mathbf{I}$ if the incident field is incoherent and maximally mixed. 
Defining the work operator
$\hat{A} = -i\omega^{-1}\oint_C \hat{S}^\dagger\,d\hat{S}$
as the loop integral of the GWS operator, the ensemble-averaged optical work produced by this state is 
\begin{equation}
  \langle W \rangle 
  = \operatorname{Tr}(\hat{A}\,\hat{\rho}_0). 
  \label{eq:Waver}
\end{equation}
For a coherent state, this equation reduces to Eq.~(2) of the main text. For a maximally mixed state,  because $\operatorname{Tr}(\hat{A}) = 0$ [see Eq.~(\ref{eq:traceless})], 
Eq.~\eqref{eq:Waver} gives
$\langle W\rangle = (P_{\rm in}/N)\operatorname{Tr}(\hat{A}) = 0$. 

\textit{Optimal wavefront shaping.---}We describe a lossless wavefront shaper by a
unitary transformation $\hat{U}$. The coherence matrix after the shaper is 
$\hat{\rho}= \hat{U}\hat{\rho}_0 \hat{U}^\dagger$. The work becomes
$\operatorname{Tr}(\hat{A}\hat{U}\hat{\rho}_0 \hat{U}^\dagger)$,
and we seek the $U$ that maximizes it. To this end, consider two eigen-decompositions: $\hat{A} = \hat{V}\hat{D}_A\hat{V}^\dagger$ with eigenvalues $\{a_k^\downarrow\}$ in decreasing orders and the corresponding eigenvectors  packed in  $\hat{V}$, as well as $\hat{\rho}_0 = \hat{Y}\hat{D}_\rho \hat{Y}^\dagger$ with eigenvalues $\{\lambda_k^\downarrow\}$ in decreasing orders and the corresponding eigenvectors packed in $\hat{Y}$. Von Neumann trace inequality~\cite{bhatiaMatrixAnalysis2013} tells us that
\begin{equation}
  W_{\rm max} \equiv \max_{\hat{U}}\,\operatorname{Tr}(\hat{A}\,\hat{U}\hat{\rho}_0\hat{U}^\dagger)
  = \sum_k a_k^\downarrow\,\lambda_k^\downarrow.
  \label{eq:Wmax}
\end{equation}
The bound is achieved by $\hat{U}_{\rm opt} = \hat{V}\hat{Y}^\dagger$, which is the optimal shaper that pairs the most powerful input mode with the
largest-eigenvalue mode of $\hat{A}$.

\textit{Monotonicity with coherence.---}
We say an incident field is more coherent than another when its coherence matrix, after normalization $\tilde\lambda_k^{(j)\downarrow}=\lambda_k^{(j)\downarrow}/\sum_{k=1}^N \lambda_k^{(j)\downarrow}$, has a more concentrated spectrum, i.e., 
\begin{equation}
\sum_{k=1}^n\tilde{\lambda}_k^{(1)\downarrow}\geq\sum_{k=1}^n\tilde{\lambda}_k^{(2)\downarrow}  \text{\ for all\ } n=1,2,...,N.
\label{eq:major}
\end{equation}
This is known mathematically as a majorization order~\cite{guoTransportMeasurementsMajorization2025}.
This characterization of coherence is suitable for the expression of optical work in Eq.~(\ref{eq:Wmax}) because the expression itself is Schur-convex, meaning if $\tilde{\lambda}_k^{(1)\downarrow}$  and $\tilde{\lambda}_k^{(2)\downarrow}$ satisfy Eq.~(\ref{eq:major}), then $W_{\rm max}(\tilde{\lambda}_k^{(1)\downarrow})\geq W_{\rm max}(\tilde{\lambda}_k^{(2)\downarrow})$.
Physically, this means the more coherent the incident field is, the more optical work we can expect to achieve via wavefront shaping.  In particular, fully coherent
illumination ($\tilde\lambda_k=\delta_{1k}$) gives the global maximum
$a_1^\downarrow$, while the maximally incoherent illumination ($\tilde\lambda_k=1/N$ for all $k$) always gives zero optical work. 

The same
argument applies to any observable quadratic function in the input field with a Hermitian
response matrix; see
Refs.~\cite{guoTransportMeasurementsMajorization2025} for a related majorization analysis of transport observables for the scattering of partially coherent waves.
}

%\onecolumngrid

\end{document}